# Compatibility of the theory of special relativity with an absolute reference frame with a longitudinal Doppler shift

Masanori Sato

*Honda Electronics Co., Ltd.,*

*20 Oyamazuka, Oiwa-cho, Toyohashi, Aichi 441-3193, Japan*

**Abstract:** We show the compatibility of the theory of special relativity with the absolute reference frame with a longitudinal Doppler shift. Using two absolute velocities $v_A$ and $v_S$ ($v_A \geq v_S$), the relative velocity $u$ is derived as $u = \dfrac{v_A - v_S}{1 - \dfrac{v_S v_A}{c^2}}$. Then, the Doppler frequency is derived as $f_0 \sqrt{\dfrac{1 - \dfrac{u}{c}}{1 + \dfrac{u}{c}}}$ using the relative velocity $u$. We also show the method for detecting the absolute reference frame. The representation of the theory of special relativity using the absolute reference frame appears more intuitive than an orthodox interpretation.

1. Introduction

Up to now, there have been no proofs of the incompatibility of the theory of special relativity with the absolute reference frame (i.e., the reference frame at rest): that is, the absolute reference frame has not been theoretically or experimentally denied.

Sato [1, 2] argued that the Michelson-Morley interference experiment cannot detect earth drift in the absolute reference frame. This is because the Michelson-Morley experiment showed an interference condition and did not show the simultaneous arrival of two photons. This is because the single photon Michelson-Morley interference experiment, as shown in **Fig. 1**, can be carried out only where there is a single photon. It was pointed out that the Michelson-Morley interference experiment cannot detect any experimental setup motion in the absolute reference frame.

Counterintuitive aspects of the theory of special relativity have arisen because the absolute reference frame is not adopted. If the absolute reference frame in the theory of special relativity is discussed clearly, a more intuitive interpretation of the theory of special relativity is possible. For example, the twin paradox does not appear because the trajectories of the twins are clearly identified not only in the absolute reference frame but also in absolute velocity space.

The longitudinal Doppler shift between two moving bodies in the absolute reference frame has already been derived [3]. In this paper, we show the compatibility of the theory of special relativity with the absolute reference frame in this longitudinal Doppler shift. Furthermore, we show how to find a stationary state in the absolute reference frame.

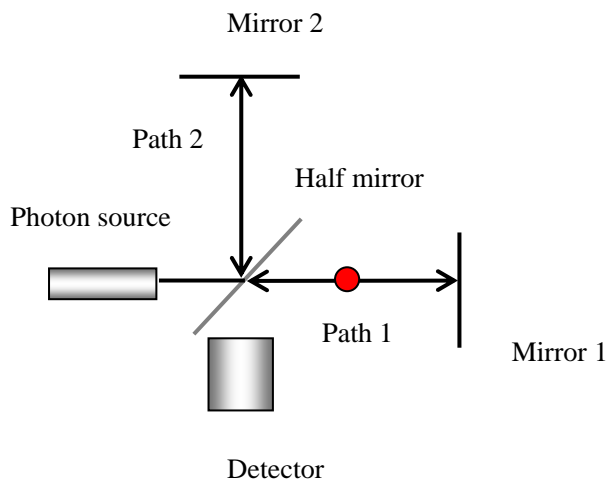

**Fig. 1** Conceptual diagram of single photon Michelson-Morley experiment: there is only a single photon in the interferometer, then the arrival time between paths 1 and 2 cannot be measured. This experiment shows that the interference condition does not depend on the earth motion. That is, we cannot detect any earth motion using the Michelson interferometer.

2  Representation of relative velocity

The relative velocity of rockets S and A in **Fig. 2** is derived. Where $v_S$ is the absolute velocity of rocket S and $v_A$ is that of rocket A, and $v_A \geq v_S$. It is rather difficult to derive the equation of relative velocity. From the analogy of the relativistic velocity addition law, we do not predict the relative velocity as $v = v_A - v_S$.

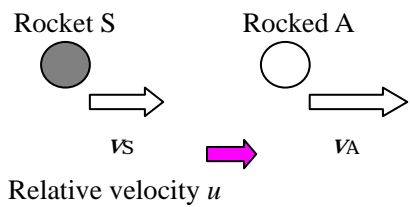

**Fig. 2** Absolute velocity of rockets S and A: If the absolute velocities of rockets S and A are assumed as $v_S$ and $v_A$ respectively, the relative velocity of rockets S and A is calculated by equation (1).

If the relativistic velocity addition law is tentatively used to obtain the relative velocity, where the two velocities are $-v_S$ and $v_A$, the relative velocity between rockets S and A is represented as follows[*A],

$$u = \frac{-v_S + v_A}{1 + \frac{(-v_S)v_A}{c^2}} = \frac{v_A - v_S}{1 - \frac{v_S v_A}{c^2}}, \tag{1}$$

where $c$ is the speed of light. The equation for the longitudinal Doppler shift frequency using absolute velocities was derived as follows [3],

$$f_{S \to A}^{L} = f_{A \to S}^{L} = f_0 \sqrt{\frac{\left(1 - \frac{v_A}{c}\right)\left(1 + \frac{v_S}{c}\right)}{\left(1 - \frac{v_S}{c}\right)\left(1 + \frac{v_A}{c}\right)}}. \tag{2}$$

Where $f_{S \to A}^{L}$ is the longitudinal Doppler shift frequency, subscript $S \to A$ denotes that rocket S sees rocket A ($A \to S$ denotes rocket A sees rocket S), and $f_0$ is the reference frequency in the reference frame at rest [3]. Thus, the equation for the Doppler shift is rewritten using equation (1) as

$$f_{S \to A}^{L} = f_{A \to S}^{L} = f_0 \sqrt{\frac{1 - \frac{v_A - v_S}{c\left(1 - \frac{v_S v_A}{c^2}\right)}}{1 + \frac{v_A - v_S}{c\left(1 - \frac{v_S v_A}{c^2}\right)}}} = f_0 \sqrt{\frac{1 - \frac{u}{c}}{1 + \frac{u}{c}}}. \tag{3}$$

The velocity $u$ in equation (3) contains absolute velocities $v_S$ and $v_A$ as represented in equation (1). However, equation (3) is in good agreement with the orthodox representation of the Doppler shift frequency. Through modification of equation (1), the postulate of an absolute reference frame is compatible with the essence of the Doppler shift representation in the theory of special relativity. The essence is that the Doppler shift frequency can be represented by the form of equation (3): that is, it depends only on the relative velocity $u$.

3. Application to numerical calculation

If we assume that the two rockets are traveling at speeds of 40% (rocket S) and 60% (rocket A) of the speed of light $c$ as shown in **Fig. 2** then according to equation (3) we obtain the relative velocity between rockets S and A as

$$u = \frac{0.6c - 0.4c}{1 - \frac{0.6 \times 0.4 c^2}{c^2}} = \frac{0.2}{0.76}c = \frac{5}{19}c.$$

This velocity $u$ shows not only the velocity that rocket S sees rockets A but also that at which rocket A sees rocket S, thus we can calculate the Doppler shift frequency using equation (3): that is the essence of the orthodox representation of Doppler shift.

Let us consider another rocket A' that is traveling slower than rocket S with the relative velocity of $\frac{5}{19}c$ as shown in **Fig. 3**. The absolute velocity of rocket A' ($v_{A'}$) can be calculated. The velocity that rocket S sees rocket A' is set to equal to the relative velocity between rocket S and rocket A ($=\frac{5}{19}c$).

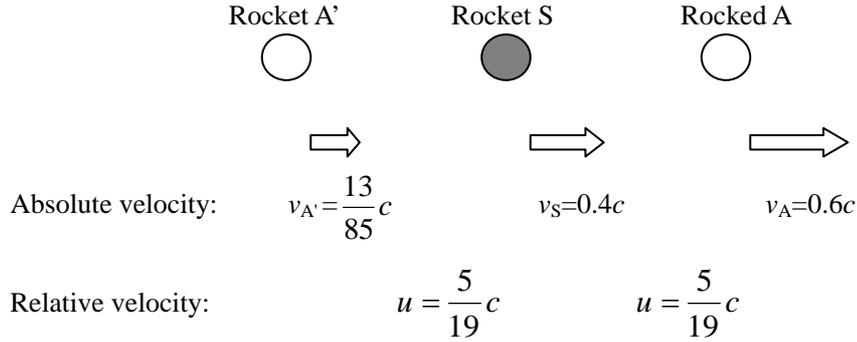

**Fig. 3** Absolute velocities of rockets S, A, and A' are summarized in Table 1. The relative velocity between rocket S and rockets A and A' is $u=\frac{5}{19}c$: that is, the relative velocities are the same and the four Doppler frequencies are the same as shown in Table 2.

Thus, we calculate $v_{A'}$ using equation (1) as,

$$u = \frac{0.4c - v_{A'}}{1 - \frac{v_{A'} \times 0.4c}{c^2}} = \frac{5}{19}c.$$

Then, we obtain $v_{A'} = \frac{13}{85}c$.

The absolute velocities of rocket S, A, and A' are shown in Table 1. The relative velocities and the Doppler frequencies between rocket S and rockets A and A' are calculated using equation (1). The values in Table 2 show the essence of the orthodox Doppler shift, that is

$$f^L_{S \to A} = f^L_{A \to S} = f^L_{S \to A'} = f^L_{A' \to S} = f_0 \sqrt{\frac{7}{12}}. \tag{4}$$

Table 1  Absolute velocities of rockets S, A, and A' in **Fig. 3**

|  | Rocket A' | Rocket S | Rocked A |
|---|---|---|---|
| Absolute velocity | $v_{A'} = \frac{13}{85}c$ | $v_S=0.4c$ | $v_A=0.6c$ |

Table 2  Relative velocities and Doppler frequencies of rockets S, A, and A' in **Fig. 3**

|  | S sees A | A sees S | S sees A' | A' sees S |
|---|---|---|---|---|
| Relative velocity | $u = \dfrac{5}{19}c$ | $u = \dfrac{5}{19}c$ | $u = \dfrac{5}{19}c$ | $u = \dfrac{5}{19}c$ |
| Doppler frequency | $f^L_{S \to A} = f_0 \sqrt{\dfrac{7}{12}}$ | $f^L_{A \to S} = f_0 \sqrt{\dfrac{7}{12}}$ | $f^L_{S \to A'} = f_0 \sqrt{\dfrac{7}{12}}$ | $f^L_{A' \to S} = f_0 \sqrt{\dfrac{7}{12}}$ |

4. To find the reference frame at rest using the sing around method

In section 3, we showed that the postulate of the absolute reference frame is compatible with the essence of the Doppler shift representation in the theory of special relativity. That is the Doppler shift frequency depends on the relative velocity which is derived using the relativistic velocity addition law. We also showed that the relative velocity can be experimentally detected using the Doppler shift frequency.

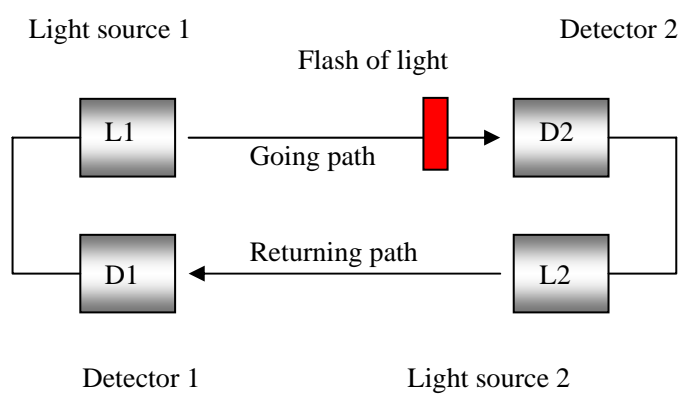

**Fig. 4**  Sing around experimental setup using light

A flash of light from light source 1 is detected by detector 2. After its detection, a new flash of light is transmitted by light source 2, detected by detector 1, transmitted again by light source 1, and so on. This experiment is modern version of Galileo's experiment using lanterns on top of two mountains shown in **Fig. 5**.

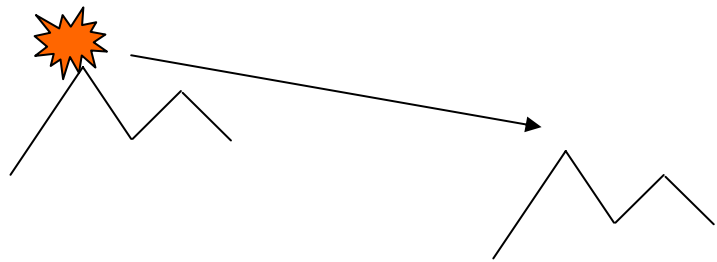

**Fig. 5**  Galileo's experiment using lanterns on top of two mountains.
They attempted to measure the traveling time of the flash of lanterns.

Table 3   Comparison of velocity measurement

| Measurements \ Effects | Interference | Lorentz transformation | | Effect of distance difference between light source and detector |
|---|---|---|---|---|
| | | light source | Reference time of detector | |
| Doppler shift | No | **Yes** | Yes | Yes |
| Sing around | No | **No** | Yes | Yes |
| Michelson-Morley interference experiment | Yes | (No) | (No) | - |

This section describes a simple method which can be used to find the reference frame at rest. Instead of the Doppler shift, the sing around method, which was previously attempted by Galileo using lanterns on top of two mountains (**Fig. 5**), is applied. The sing around experimental setup using a light source [1, 3], which uses two pairs of light sources and detectors as shown in **Fig. 4**, can be constructed, where a pulsed signal (a flash of light) is transmitted by light source 1 and detected by detector 2. After its detection, a new pulsed signal is transmitted by light source 2, detected by detector 1, transmitted again by light source 1, and so on. Light sources are constructed using LEDs, and the detection is made by photo diodes. We can measure the repetition frequency of the flash of light.

Several pairs of rockets are launched from rocket S for all directions. Let us consider two pairs of rockets, one pair consists of rockets A and A' traveling parallel to rocket S and the other consists of rockets B and B' traveling vertical to rocket S. **Figure 6** shows two pairs of rockets. The relative velocity between rockets S and A is $\frac{5}{19}c$, the relative velocity between rockets S and B is also $\frac{5}{19}c$. Thus the eight Doppler frequencies are the same as

$$f^L_{S \to A} = f^L_{A \to S} = f^L_{S \to A'} = f^L_{A' \to S} = f^L_{S \to B} = f^L_{B \to S} = f^L_{S \to B'} = f^L_{B' \to S} = f_0 \sqrt{\frac{7}{12}}. \tag{5}$$

Equation (4) indicates that the relative velocities can be experimentally defined using the Doppler shift frequencies: the rockets should be accelerated to obtain the same Doppler shift frequency. If there is another method to define the velocity that is different from the Doppler shift, we can detect the motion of rocket S in the absolute reference frame (we can calculate the absolute velocity). The Doppler shift frequency is affected by the motion of the light source and the detector: that is, the light source frequency and the reference time of detector are affected by the motion. The sing around method does not use the reference time but only compares the repletion timing of the sing around signals. For example, the repletion timings of the sing around signals of rockets A and A' is gradually shifting as shown in **Fig. 5**. On the other hand, the signals of rockets B and B' are completely coincident as shown in **Fig. 6**. Thus we can know the direction of the rocket S motion in the absolute reference frame. The direction of the motion of rocket S is toward rocket A. Thus, we know the direction to decelerate rocket S toward rocket A'. Then we can put the rocket S in a stationary state.

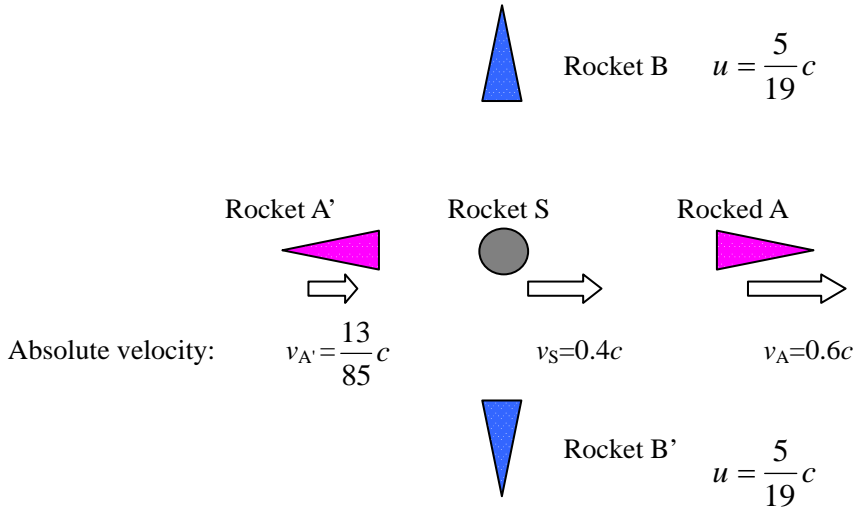

**Fig. 6** To find the reference frame at rest using the sing around method: rockets A and A' travel parallel to rocket S, rockets B and B' vertical to the motion of rocket S. The relative velocity between rocket S and rockets A, A', B, and B' is calculated as $u = \frac{5}{19}c$. The Doppler frequency is represented as equation (4). The repletion timings of sing around pulses are gradually shifted between rockets A and A' thus we know the absolute motion of rocket S.

In this discussion, only Einstein's assumption, in which a photon travels at the speed of light, $c$, in vacuum regardless of the velocity of the light source, is used; thus, the three rockets will detect the flashes of light as shown in **Fig. 5**. Rockets A and A' are launched in opposite directions from rocket S to detect the same Doppler shift frequency, the flash pattern of light that rocket S detects is shown as the points on the line $v=0.4c$ in **Fig. 5**.

The sing around method does not care the Doppler shift of the flash of light. It requests only a repetition timing of the flash of light; therefore, the discussion becomes clear and simple. We propose a new method that is different from the Michelson-Morley interference experiment to detect the motion in the absolute reference frame.

The experimental procedure to find the reference frame at rest is:

(1) Rockets are launched in all directions from rocket S and accelerated so as to detect the same Doppler shift frequency: $f^L_{S \to A} = f^L_{S \to A'} = f^L_{S \to B} = f^L_{S \to B'} = \cdots$ and so on, if the same Doppler shift frequency is detected all rockets are in the same relative velocity to rocket S.

(2) Detect the timings of sing around pulses between the pair rockets. Rockets S is moving in the

direction of the rockets pair which shows the largest timing difference of sing around light pulses: in **Fig. 8** rocket S travels toward rocket A. Then, decelerate rocket S toward rocket A'.

(3) Repeat the procedure (1) and (2) so as to rocket S detect the same timings of sing around pulses for all pairs of rockets.

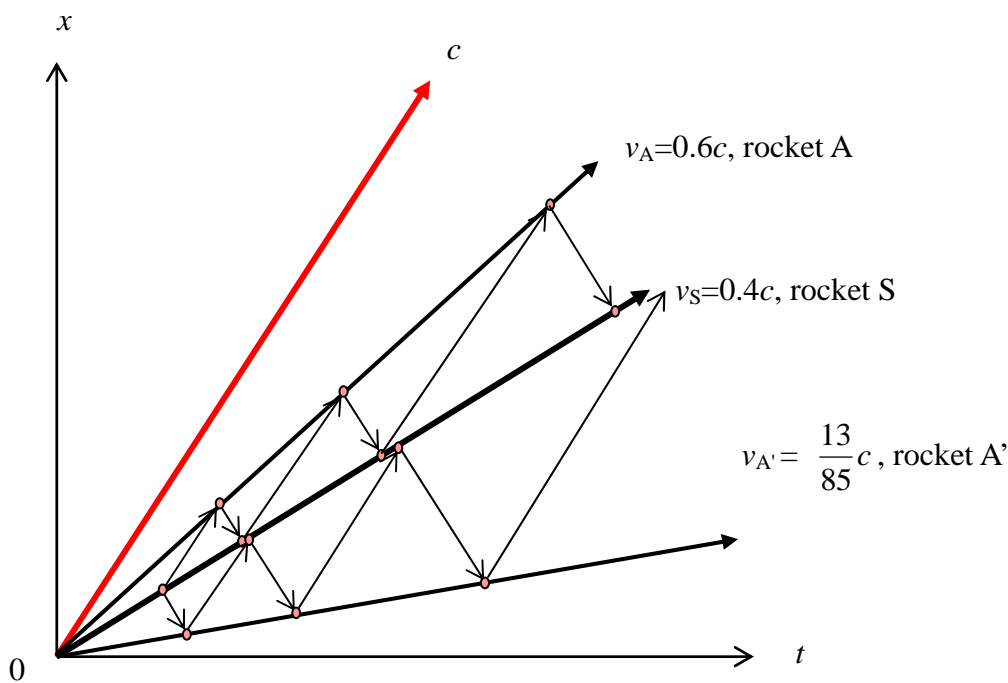

**Fig. 7** The sing around method of a light source: in this case rocket S travels toward rocket A. Rockets A' and A are launched from rocket S in opposite direction in order to detect the same Doppler shift frequency. The speed of light, $c$ is assumed to be constant regardless of the velocity of the light source. The light paths are drawn in the figure. The Doppler frequencies of rockets A and A' shown from rocket S can also be represented as $f^L_{S \to A} = f^L_{S \to A'}$. The relative velocities between rocket S and rockets A and A' are equal. However, the arrival timings of the return pulses detected at rocket S gradually shift. This indicates that rocket S is not in a stationary state in the absolute reference frame, and rocket S should be decelerated toward rocket A' which is closer to the stationary state.

We have two methods to define the velocity: one is the Doppler shift frequency and the other is the repetition timings of the sing around light pluses. Thus we can detect the reference frame at rest.

The timings shown on line $v_S=0.4c$ in **Fig, 7**, three rockets travel as shown in **Fig. 3,** thus, rocket S is decelerated to leave rocket A (accelerated toward rocket A'). **Figure 8** shows the transverse motion of rockets B and B' with rocket S. The repetition timings of sing around light pulses are coincident. **Figure 8** also represents the motion of rockets A and A' when rocket S is in a stationary state in the absolute reference frame. The condition that not only the Doppler shift frequency but also the repetition timings of sing around are equivalent indicates that rocket S is in a stationary state.

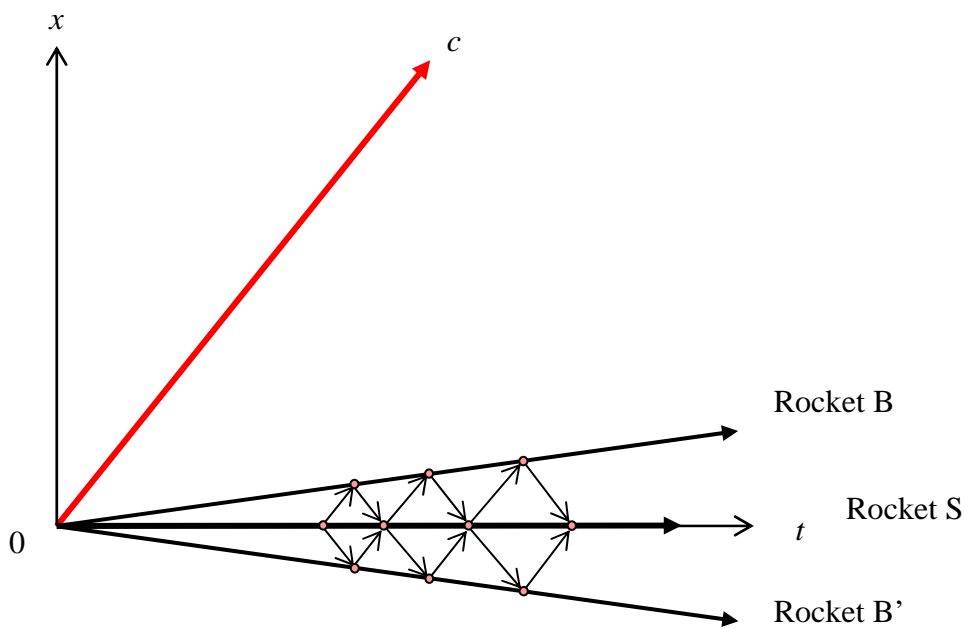

**Fig. 8** Sing around pulses from rockets B and B': in this case rockets B and B' travel vertical to rocket S, or rocket S is in the stationary state. Rocket S detects not only the same Doppler shift frequency (i.e., $f_{S \to B}^{L} = f_{B \to S}^{L} = f_{S \to B'}^{L} = f_{B' \to S}^{L}$) but also the same repetition timings of the sing around. This condition will also be detected with rockets A and A' when rocket S is in a stationary state.

5. Merit of the representation using the absolute reference frame

We showed that the absolute reference frame is compatible with the orthodox Doppler shift equation. It is said that there is no absolute reference frame: however, there is no proof of incompatibility between the theory of special relativity and the absolute reference frame. We showed that the absolute reference frame can be detected. We are able to know not only the relative velocity but also the absolute velocities of the two rockets.

If we adopt the absolute reference frame, there is no twin paradox because the traveling paths of twins can be clearly identified in the absolute reference frame. Furthermore, the paths of twins in the absolute velocity space can be identified. The merit of the representation using the absolute reference frame is that a more intuitive interpretation of the theory of special relativity is possible.

6. Conclusion

We showed the compatibility of the theory of special relativity with the absolute reference frame with a longitudinal Doppler shift. Using two absolute velocities $v_A$ and $v_S$ ($v_A \geq v_S$), the relative velocity $u$ is represented as $u = \dfrac{v_A - v_S}{1 - \dfrac{v_S v_A}{c^2}}$, thus, the Doppler frequency is derived as $f_0 \sqrt{\dfrac{1 - \dfrac{u}{c}}{1 + \dfrac{u}{c}}}$.

This representation satisfies the essence of the theory of special relativity. We also showed one way to detect the absolute reference frame. The representation of the theory of special relativity using the absolute reference frame makes a more intuitive interpretation possible.

At this stage the postulate of the absolute reference frame can be only discussed as a thought experiment because there is no experimental data: however this postulate can prevent the twin paradox from entering the theory of special relativity without any modification to the theory of special relativity.

References

1) M. Sato, "Proposal of Michelson-Morley experiment via single photon interferometer: Interpretation of Michelson-Morley experimental results using de Broglie-Bohm picture," physics/0411217, (2004).
2) M. Sato, "Interpretation of the slight periodic displacement in the Michelson-Morley experiments," physics/0605067, (2006)
3) M. Sato, "Derivation of longitudinal Doppler shift equation between two moving bodies in a reference frame at rest using the particle property of photons," physics/0502151, (2005).

\* Appendix A

The relative velocity addition in **Fig. 1**, where $v_A$ is the summation of $v_S$ and $u$, is represented as,

$$v_A = \frac{v_S + u}{1 + \dfrac{v_S u}{c^2}}. \tag{A-1}$$

Thus, the relative velocity $u$ is derived as,

$$u = \frac{v_A - v_S}{1 - \dfrac{v_S v_A}{c^2}}. \tag{1}$$

Then using equation (1), the Doppler frequency between rockets S and A is represented as,

$$f_{S \to A}^{L} = f_{A \to S}^{L} = f_0 \sqrt{\frac{1 - \frac{u}{c}}{1 + \frac{u}{c}}}. \tag{3}$$

Equation (3) can be rewritten as,

$$f_0 \sqrt{\frac{1 - \frac{u}{c}}{1 + \frac{u}{c}}} = f_0 \sqrt{\frac{1 - \frac{v_A - v_S}{c\left(1 - \frac{v_S v_A}{c^2}\right)}}{1 + \frac{v_A - v_S}{c\left(1 - \frac{v_S v_A}{c^2}\right)}}} = f_0 \sqrt{\frac{\left(1 - \frac{v_A}{c}\right)\left(1 + \frac{v_S}{c}\right)}{\left(1 - \frac{v_S}{c}\right)\left(1 + \frac{v_A}{c}\right)}}. \tag{2}$$

Equation (2) shows the Doppler frequency between two bodies with the absolute velocities of $v_A$ and $v_S$ [3].